\documentclass{appolb}
\usepackage{graphicx}
\usepackage{color}
\usepackage{hyperref}
\usepackage{xcolor}
\usepackage{lineno}
\usepackage{ulem}

\newcommand{\km}{K^{\!^-}\!\!({ \bar{u}} { s})}
\newcommand{\ph}{\phi({s} {\bar{s}})}
\newcommand{\al}{\bar{\Lambda}(\bar{u}\bar{d}\bar{s})}
\newcommand{\pbar}{\bar{p}({ \bar{u}\bar{u}}{ \bar{d}})}
\newcommand{\ks}{\overline{\Xi}^{+}(\bar{d}\bar{s}\bar{s})}
\newcommand{\om}{{\Omega}^{-}(sss)}
\newcommand{\op}{\overline{\Omega}^{+}(\bar{s}\bar{s}\bar{s})}

\begin{document}
\title{The splitting of directed flow for identified light hadrons ($K$ and $p$) and strange baryons ($\Xi$ and $\Omega$) in Au+Au collisions at STAR 
\thanks{Presented at 29th International Conference on Ultra-relativistic Nucleus-Nucleus Collisions (Quark Matter 2022), Krakow, Poland }%
}
\author{Ashik Ikbal Sheikh (for the STAR Collaboration)
\address{Department of Physics, Kent State University,\\
Kent, OH 44242, USA\\
Email: asheikh2@kent.edu, ashikhep@gmail.com}
}
\maketitle
\begin{abstract}
The first measurements for rapidity-odd directed flow of $\Xi$ and $\Omega$ in Au+Au collisions at $\sqrt{s_{\mathrm {NN}}}=$ 27 and 200 GeV are reported. The coalescence sum rule is examined with various combinations of hadrons where all constituent quarks are produced, such as $\km$, $\pbar$, $\al$, $\ph$, $\ks$, $\om$, and $\op$. For such combinations, a systematic violation of the sum rule is observed with increasing difference in the electric charge and the strangeness content of the combinations. Measurements are compared with the calculations of A Multi-Phase Transport (AMPT) model and Parton-Hadron String Dynamics (PHSD) model with electromagnetic (EM) field. The PHSD model with EM field agrees with the measurements within uncertainties. 
\end{abstract}
  
\section{Introduction}
Directed flow ($v_1$) is the first harmonic coefficient in the Fourier expansion of the final-state azimuthal distribution relative to the reaction plane. Theoretical calculations~\cite{Heinz:2009xj} based on nuclear transport and hydrodynamics indicate that the $v_1$ is sensitive to the early stages of the high energy heavy-ion collisions. 

One of the most important features of the early stages of the heavy-ion collisions is the production of an extremely strong magnetic field due to the motion of the charged spectators. 
The produced magnetic field decays down fast since the charged spectators fly away from the collision zone and this generates an electric current in the plasma due to the Faraday effect. In addition, the plasma has a longitudinal expansion velocity along the beam direction and perpendicular to magnetic field direction, hence the Lorentz force pushes charged particles and anti-particles of the plasma in opposite ways which are perpendicular to both the directions of longitudinal velocity of plasma and the magnetic field. This is analogous to Hall effect. The charged spectators can also generate an electric current in the plasma due to Coulomb effect. The $v_1$ of different produced charged particles is greatly influenced by the resultant of Faraday, Hall, and Coulomb effects which eventually leads to the splitting of $v_1$ ~\cite{Gursoy:2018yai}. Both STAR and ALICE experiments measured a non-zero $v_1$ splitting with pseudo-rapidity between positively and negatively charged hadrons in Au+Au collisions at $\sqrt{s_{\mathrm {NN}}}= $ 200 GeV~\cite{Adamczyk:2016eux} and Pb+Pb collisions at $\sqrt{s_{\mathrm {NN}}}=$ 2.76 TeV~\cite{Acharya:2019ijj}, respectively. However, the interpretation of the measured splitting using charged hadrons runs into difficulties especially when the effects of electromagnetic fields are concerned. This is due to the fact that among light hadrons, there are many (anti-)particles containing $u$ and $d$ quarks, which can be either transported from beam rapidity~\cite{Dunlop:2011cf} or produced in the collisions. Due to the different number of interactions suffered, the transported $u$ and $d$ have different $v_1$ than those of the produced quarks~\cite{Guo:2012qi}. This ensures a pre-existing splitting due to the transport. This transport driven splitting is a background in search of the pure electromagnetic-field-driven splitting. We could avoid the transported quarks in our analysis by selecting particles composed of produced constituent quarks only ($\bar{u}$, $\bar{d}$, $s$, and $\bar{s}$). The experimental method is outlined briefly in Sec.~\ref{approach}.

In this contribution, we report the measurements of $v_1$ of multi-strange baryons ($\Xi$ and $\Omega$).  The $v_1$-splitting as a function of electric charge difference ($\Delta q$) and strangeness difference ($\Delta S$) is measured, using $K^-$, $\bar{p}$, $\bar{\Lambda}$, $\phi$, $\overline{\Xi}^{+}$, $\Omega^{-}$, and $\overline{\Omega}^{+}$ from Au+Au collisions at $\sqrt{s_{\mathrm {NN}}}=27$ and $200$ GeV.


\section{Data sets and Analysis strategy}
\label{approach}
STAR detector system is versatile experimental setup for track reconstruction, vertexing, and particle identification at RHIC. The main sub-detectors are (i) Time Projection Chamber (TPC) ($|\eta| \le 1$): used for charged particle tracking, vertexing, and particle identification; (ii) Time-Of-Flight (TOF) detector: used for particle identification; (iii) Event-Plane Detectors (EPDs) ($2.1 < |\eta| < 5.1$) and (iv) Zero-Degree Calorimeter with Shower-Maximum Detectors (ZDC-SMDs) ($|\eta|>6.3$): can measure event planes of the collisions. 

A high statistics data samples for Au+Au collisions at$\sqrt{s_{\mathrm {NN}}}=$ 27 and 200 GeV are used in the measurements.  We use events with the vertex position along the beam direction with $|V_{\mathrm z}|<70$ cm, and along the radial directions, $V_{r}<2$ cm at $\sqrt{s_{\mathrm {NN}}}=27$ GeV; and $|V_{z}|<30$ cm, $V_{\mathrm r}<2$ cm at $\sqrt{s_{\mathrm {NN}}}=$ 200 GeV. At the track level, $p_\mathrm T\!>\!0.2$ GeV/$c$, and a distance of closest approach (DCA) from vertex, $\rm {DCA} \le 3$ cm, and at least 15 space points in the TPC acceptance are selected. 
For particle identification, we use $p_\mathrm T\!>\!0.2$ GeV/$c$, momentum $<1.6$ GeV/$c$ and $|n_{\sigma}|\le 2$ for charged pions and kaons; $\!0.4<p_\mathrm{T}\!<\!5$ GeV/$c$, $|n_{\sigma}|\le 2$ for $p$ and $\bar{p}$, where $n_{\sigma}$ is the standard deviation of difference between measured $\langle dE/dx\rangle$ and theoretical mean value for each particle type. The $\Lambda$, $\bar{\Lambda}$, ${\Xi}^{-}$, $\overline{\Xi}^{+}$, ${\Omega}^{-}$, and $\overline{\Omega}^{+}$ are reconstructed using KF-Particle package~\cite{Gorbunov:2013phd}. The $\phi$-mesons are reconstructed in $K^+K^-$ channel using the invariant mass method with pair rotation background subtraction. The systematic uncertainties on the measurements are obtained by varying these analysis cuts. We remove the effect of the statistical fluctuations by employing Barlow's method~\cite{Barlow:2002yb}.

The analysis method is based on the quark coalescence mechanism. Coalescence sum rule states that the directed flow of a hadron is consistent with the sum of the directed flow of its constituent quarks, $\it{i.e.}$, $v_1({\rm hadron}) = \sum\limits_i v_1(q_i)$, where the sum runs over the $v_1$ of the constituent quarks, $q_i$. 
There are many hadron species composed of constituent $u$ and $d$ quarks, which might or might not be transported from the incoming nuclei. The transported quarks produce background in search of the possible electromagnetic-field-driven $v_1$ splitting. Hence, in the analysis we take particles which contain produced quarks only, namely, $\km$, $\pbar$, $\al$, $\ph$, $\ks$, $\om$, and $\op$. All these particles have different flavour, electric charge ($q$) and mass ($m$); and $v_1$ is sensitive to quark flavour and mass. Keeping this in mind, we combine the different particles so that combinations have same or similar mass at the constituent quark level ($\Delta m \approx 0$), but $\Delta q \ne 0$ and $\Delta S \ne 0$. We found five independent combinations as shown in Table~\ref{tab:delq_dels}. The difference $\Delta v_1$ of such combinations is called ``splitting of $v_1$"~\cite{Sheikh:2021rew} and the slope of the $\Delta v_1$ vs. rapidity is a measure of the splitting. 
We measure the $\Delta v_1$ of all the indices in Table~\ref{tab:delq_dels} to obtain the splitting as a function of $\Delta q$, a measure of EM-field-driven splitting. At the same time, we also measure the the $\Delta v_1$ with $\Delta S$ of all the combinations. Although, the change in $\Delta q$ is also associated with a change in $\Delta S$ in Table~\ref{tab:delq_dels} which comes from the quantum numbers carried by the constituent quarks.

\begin{table*}[th]
\centering
\renewcommand{\arraystretch}{1.5}
\begin{tabular}{|l l l l l l l|}
\hline
Index & Quark mass && Charge & & Strangeness & ~~~~~$\Delta v_1$ combination   \\ 
\hline
1 & $\Delta m=0$ & & $\Delta q=0$ & & $\Delta S=0$ & ${[\pbar + \ph]}-[\km + \al ] $ \\ 
\hline
2 & $\Delta m\approx 0$ & & $\Delta q=1$ & & $\Delta S=2$ &
${[\al]}-[\frac{1}{3}\om + \frac{2}{3}\pbar]$\\
\hline
3 & $\Delta m\approx 0$ & & $\Delta q=\frac{4}{3}$ & & $\Delta S=2$ &
${[\al]}-[\km + \frac{1}{3}\pbar]$ \\
\hline
4 & $\Delta m= 0$ & & $\Delta q=2$ & & $\Delta S=6$ &
${[\op]}-[\om]$ \\
\hline
5 & $\Delta m\approx 0$ & & $\Delta q=\frac{7}{3}$ & & $\Delta S=4$ &
${[\ks]}-[\km + \frac{1}{3}\om]$ \\
\hline
\end{tabular}
\caption{Table showing difference in mass, charge, and strangeness between combinations formed from seven particle species composed of produced quarks only. 
}
\label{tab:delq_dels}
\end{table*}

\section{Results and Discussions}
\begin{figure}[htb]
\centerline{%
\includegraphics[width=5.4cm]{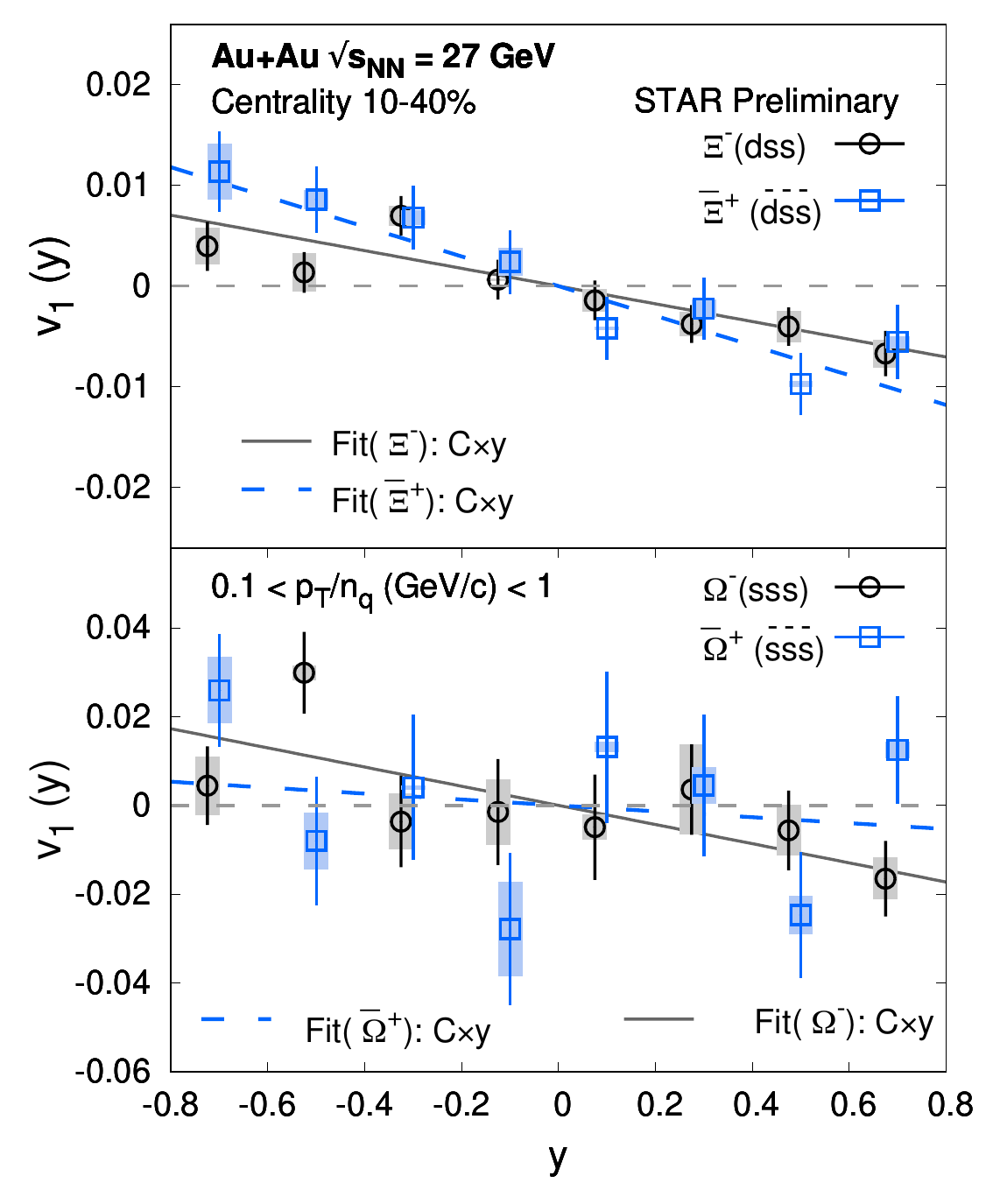}
\includegraphics[width=5.4cm]{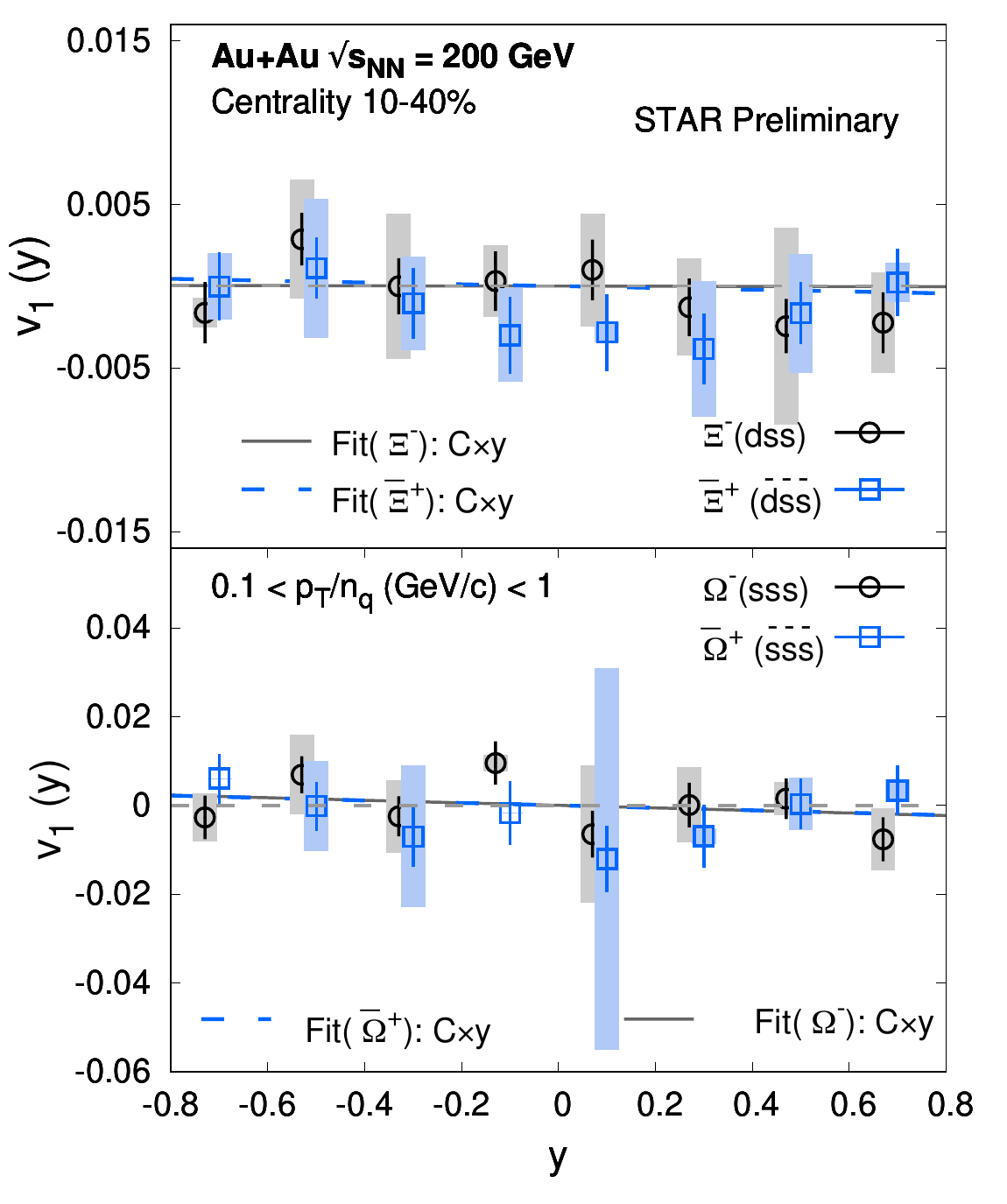}}
\caption{$v_1$ of $\Xi^{-}$, $\overline{\Xi}^{+}$, ${\Omega}^{-}$ and $\overline{\Omega}^{+}$ as a function of rapidity, $y$, in 10\%-40\% central Au+Au collisions at $\sqrt{s_{\mathrm {NN}}} = 27$ and $200$ GeV.}
\label{fig:v1_xiom}
\end{figure}

Figure~\ref{fig:v1_xiom} displays the first measurements of $\Xi$ and $\Omega$ baryon $v_1$ in 10\%-40\% central Au+Au collisions at $\sqrt {s_{NN}} = 27$ and $200$ GeV. We perform a linear fit, $v_1(y)=Cy$, where $C$ is the fitting parameter, and $y$ is the rapidity. We found: $C=-0.0083\pm0.0020~(\rm{stat.}\pm0.00~(\rm{syst.}))$ $[-0.0148\pm0.0028~(\rm{stat.})\pm0.0013~(\rm{syst.})]$ for $\Xi^{-}$ [$\overline{\Xi}^{+}$] and $C=-0.0214\pm0.008~(\rm{stat.})\pm0.0034~(\rm{syst.})$ $[-0.0075\pm0.0118~(\rm{stat.})\pm0.0017~(\rm{syst.})]$ for $\Omega^{-}$ [$\overline{\Omega}^{+}$] at $\sqrt {s_{NN}} = 27$ GeV. There is a hint of a larger $v_1$ for ${\Omega}^{-}$ compared to $\Xi$ baryons is observed at $\sqrt{s_{\mathrm {NN}}} = 27$ GeV, though the uncertainties are large. 

\begin{figure}[htb]
\centerline{%
\includegraphics[width=7.1cm]{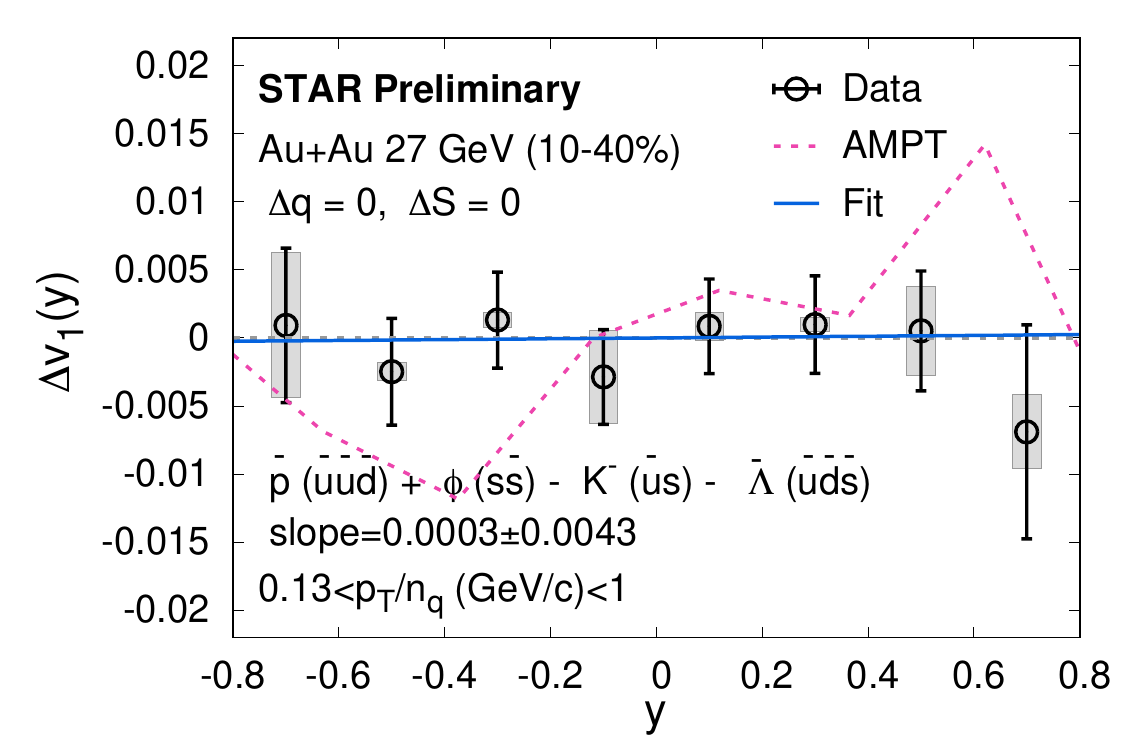}
\includegraphics[width=7.1cm]{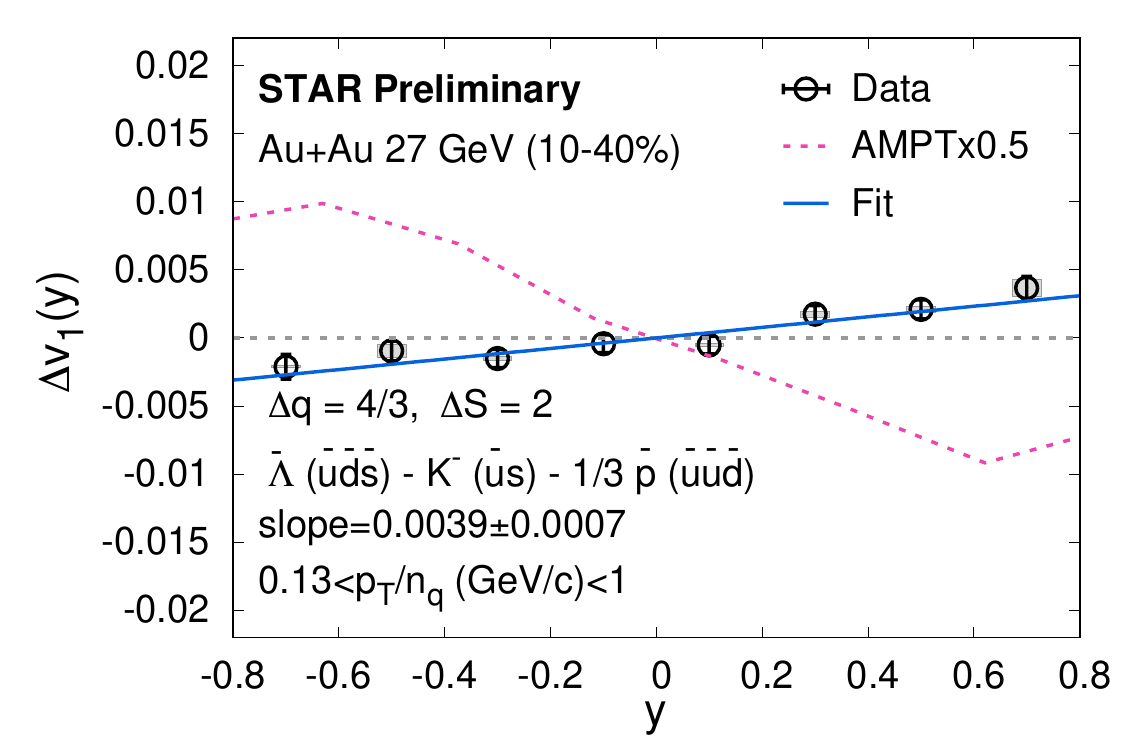}}
\caption{$\Delta v_1$ as a function of $y$ for ($\Delta q$, $\Delta S$) $=$ (0, 0), (4/3, 2) in Au+Au collisions at $\sqrt{s_{\mathrm {NN}}}=27$ GeV in 10\%-40\% centrality.}
\label{fig:delv1y}
\end{figure}

In Fig.~\ref{fig:delv1y}, we show the measured $\Delta v_1(y)$ for hadron combinations with ($\Delta q$, $\Delta S$) $=$ (0, 0), (4/3, 2) in 10\%-40\% Au+Au collisions at $\sqrt{s_{\mathrm {NN}}}=27$ GeV. The $\Delta v_1$-slope parameters of the measurements are extracted. For $\Delta q = 0$ and $\Delta S = 0$ (identical quark combination case), the value of the slope is a minimum compared to $\Delta q$ = 4/3 and $\Delta S$ = 2 cases. This minimum deviation from zero implies that the coalescence sum rule holds with the identical quark combination. The deviation of the slope from zero increases as we move to $\Delta q$ = 4/3 and $\Delta S$ = 2 case. A Multi-Phase Transport (AMPT)~\cite{Lin:2004en,Nayak:2019vtn} model calculation can describe the measured $\Delta v_1$ within errors for the $\Delta q = 0$, $\Delta S = 0$ case. For $\Delta q$ = 4/3 and $\Delta S$ = 2, AMPT depicts a completely opposite trend compared to the data.

\begin{figure}[htb]
\centerline{%
\includegraphics[width=8.2cm]{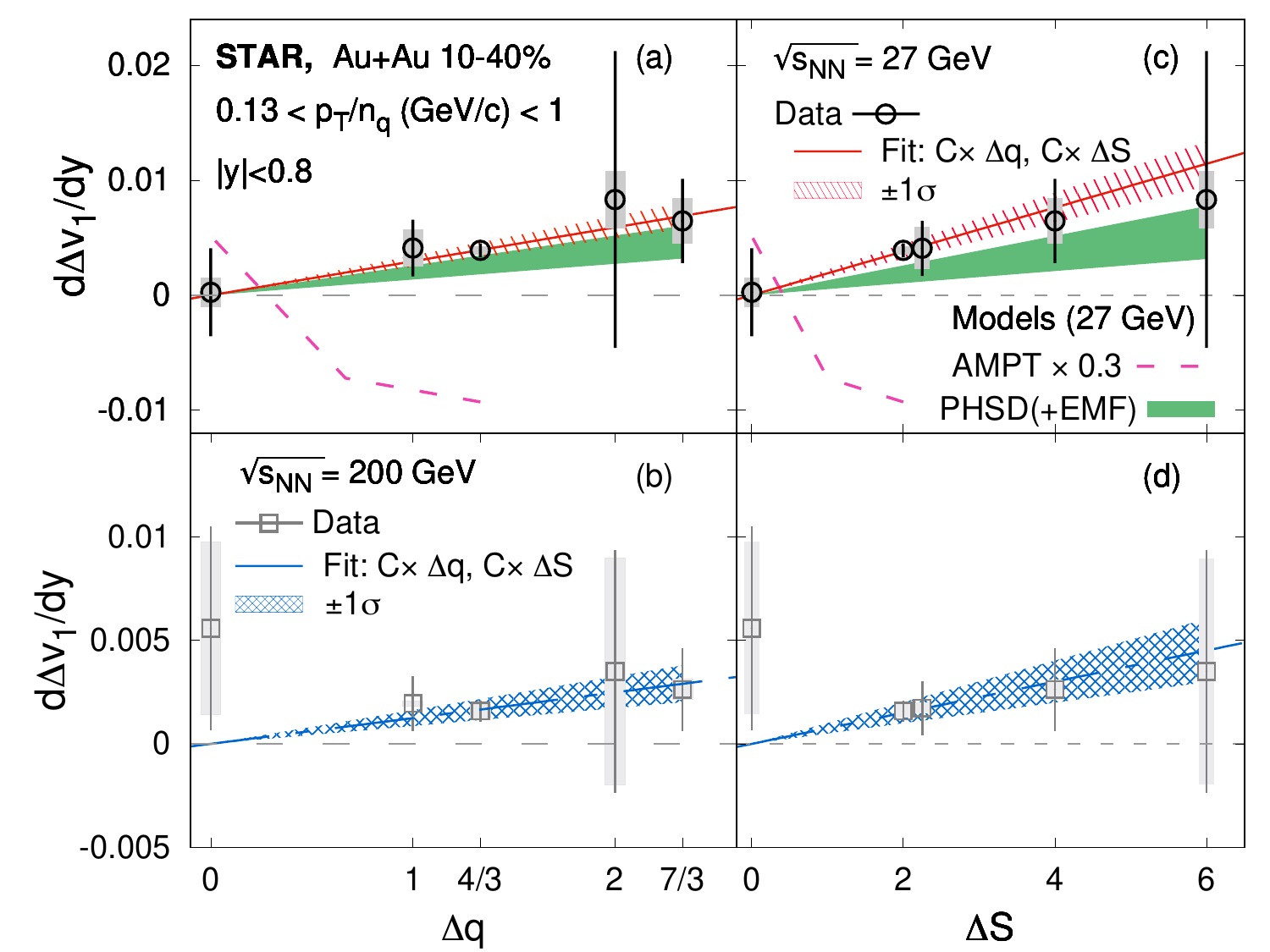}}
\caption{$\Delta v_1$-slope ($d\Delta v_1/dy$) as a function of $\Delta q$, and $\Delta S$ for 10\%-40\% centrality in Au+Au collisions at $\sqrt{s_{\mathrm {NN}}}=27$ GeV and $\sqrt{s_{\mathrm {NN}}}=27$ GeV.}
\label{fig:slope_delq_dels}
\end{figure}

In Fig.~\ref{fig:slope_delq_dels}, we display the mid-rapidity $\Delta v_1$-slope ($d\Delta v_1/dy$) as a function of $\Delta q$ and $\Delta S$ for 10\%-40\% central Au+Au collisions at $\sqrt{s_{\mathrm {NN}}}=27$ and 200 GeV. The $d\Delta v_1/dy$ increases with $\Delta q$ and $\Delta S$. The slope parameters of the $d\Delta v_1/dy$ with $\Delta q$, $d^2\Delta v_1/dy\,d\Delta q$, are $[2.952\pm0.489~(\rm{stat.})\pm0.367~(\rm{syst.})]\times10^{-3}$ and $[1.242\pm0.381~(\rm{stat.})\pm0.258~(\rm{syst.})]\times10^{-3}$ at $\sqrt{s_{\mathrm {NN}}}=27$ and 200 GeV, respectively.
 The magnitude of the $\Delta v_1$ slope is larger at $\sqrt{s_{\mathrm {NN}}}=27$ GeV than at 200 GeV with $4.83\sigma$ significance. The AMPT calculations~\cite{Lin:2004en,Nayak:2019vtn} do not agree with the measurements whereas the PHSD with EM field calculations can explain the data within uncertainties. The PHSD model with EM field assumes that all electric charges are affected by the strong EM field which ensures splitting of $v_1$ between positive and negative particles as observed in Fig~\ref{fig:slope_delq_dels}.

\section{Summary}

In summary, we present the first measurements of directed flow, $v_1(y)$, of $\Xi$ and $\Omega$ in Au+Au collisions at $\sqrt{s_{\mathrm {NN}}}=27$ GeV and 200 GeV. There is a hint of a relatively larger $v_1$-slope for $\Omega^{-}$ compared to the $\Xi$ baryons within the uncertainties. We measure directed flow splitting, $\Delta v_1$, with $\Delta q$ and $\Delta S$. The $\Delta v_1$-slope for hadron combinations, increases with $\Delta q$ and $\Delta S$. The strength of the splitting increases going from $\sqrt{s_{\mathrm {NN}}}=$ 200 to 27 GeV. The PHSD with EM field calculations can describe the $\Delta q$ and $\Delta S$ dependent splitting within uncertainties. 

\section*{Acknowledgments}
Author acknowledges support from the Office of Nuclear Physics within the US DOE Office of Science, under Grant DE-FG02-89ER40531.

\end{document}